\documentclass[letterpaper]{jpconf}
\usepackage[dvips]{graphicx,epsfig,color}
\begin{document}
\title{Search for $ZH$ Production at D0 in $p\bar{p}\to\ell^+\ell^-b\bar{b}$ Events at $\sqrt{s}=1.96$ TeV}

\author{John BackusMayes on behalf of the D\O\ collaboration}

\address{University of Washington}

\begin{abstract}
We present a search for a low-mass standard model Higgs boson produced in association
with a $Z$ boson decaying to charged leptons at a center-of-mass energy of $\sqrt{s}=$1.96~TeV
with the D0 detector at the Fermilab Tevatron collider. The search is performed
in a large data set of events containing two opposite-sign leptons (electron, muon, tau)
and one or two b-tagged jets. Recent improvements to the sensitivity, from increased lepton
acceptance to optimized signal-to-background discrimination, will be discussed.  This article
is part of the conference proceedings for the 2010 Lake Louise Winter Institute.
\end{abstract}

\section{Introduction}

The mass of a standard model Higgs boson is constrained by direct searches
performed at LEP and measurements of the top quark and W boson masses \cite{ewwg}.
Combining these results, the Higgs mass $m_H$ must be less than 186 GeV at 95\% confidence level.

If $m_H < 135$ GeV, then Higgs bosons are expected to decay primarily to $b\bar{b}$.
At hadron colliders, the inclusive $b\bar{b}$ cross section is six orders of magnitude
larger than the cross section for Higgs production, so it is not feasible
to find evidence for low-mass Higgs bosons produced alone. Instead, we search for
associated production of vector ($W$, $Z$) and Higgs bosons. Requiring leptonic decay of the $W$ or $Z$
dramatically reduces the multijet background and increases our sensitivity to the Higgs signal.
The analysis discussed here is concerned exclusively with $ZH$ production in the $\ell^+\ell^-b\bar{b}$
final state; the D0 collaboration has also completed similar analyses using $\ell\nu b\bar{b}$ \cite{lnubb}
and $\nu\bar{\nu}b\bar{b}$ \cite{nunubb} final states.

The irreducible backgrounds in this search are $Z$ production with heavy-flavor jets,
top quark pair production, and diboson final states. Instrumental backgrounds include
jets faking charged leptons and light jets faking heavy-flavor jets.

A more detailed description of this analysis may be found in \cite{moriond09}.

\section{The Tevatron and the D0 detector}

The Tevatron is a proton-antiproton collider located at Fermilab near Chicago, IL.
Collisions have a center-of-mass energy of $\sqrt{s}=1.96$ TeV.
Fermilab's Accelerator Division continues to optimize the integrated luminosity
produced by the Tevatron, and currently the accelerator is performing better than ever before.
The total integrated luminosity delivered from RunII is over 7 $\textrm{fb}^{-1}$;
results discussed here use up to 4.1 $\textrm{fb}^{-1}$ of data.

D0 is a multi-purpose particle detector, one of two detectors located at collision points
around the Tevatron. We have taken data with 90\% average efficiency since the
start of RunII. In this search, we employ every major component of D0 in order to
identify muons, electrons, and heavy-flavor jets \cite{d0_detector}.

\section{Event selection}

\subsection{Leptons}

We strive for maximum Higgs signal acceptance, so our event selection is very loose.
For muons, we require central track matches, $p_T > 10$ GeV, and $|\eta| < 2$. Electron requirements are
$p_T > 15$ GeV and $|\eta| < 2.5$.  All identified leptons are also required to pass various tracking and
calorimeter isolation criteria, and the invariant mass of the dilepton pair must match the $Z$ boson
resonance: $70 < m_{\ell\ell} < 130$ GeV.

To be sure we accept as many Higgs events as possible, we select some electrons and muons
that are not initially identified as such.
In the inter-cryostat region (ICR) where there is little calorimeter coverage, we look for
electrons that have been reconstructed as taus. In the various gaps in muon coverage,
we look for isolated tracks. These additions improve our signal acceptance by 15\%.

\subsection{Jets}

We select events with at least 2 jets, leading jet $p_T > 20$ GeV and second jet $p_T > 15$ GeV.
Before tagging, $S/B = 0.0003$.

Using D0's neural net b-tagging algorithm \cite{btagnn}, we require either two loose tags (inclusive) or
one tight tag (exclusive), which improves $S/B$ by factors of 20 or 10, respectively.
The optimization of our final multivariate discriminant depends significantly on b-tag
criteria, as do the expected signal and background yields, so we can improve our Higgs sensitivity
by analyzing these orthogonal b-tag samples separately.

With the two highest-$p_T$ tagged jets or the one tagged jet and highest-$p_T$ untagged jet,
we compute the invariant mass of the dijet system, which is the kinematical variable most
sensitive to low-mass Higgs production.

\subsection{Kinematic fit}

With an ideal detector, we would have very little missing $E_T$ in $ZH\to\ell^+\ell^-b\bar{b}$ events.
Thus, events with a large $p_T$ imbalance must result from either background processes or mismeasurement.
Given our knowledge of the jet and lepton energy resolutions, we can make our measurements more precise
and discriminate against backgrounds such as multijet production and $t\bar{t}\to\ell^+\ell^-\nu\bar{\nu}b\bar{b}$.
To do this, we perform a constrained multi-dimensional fit
on the $p_T$, $\eta$, and azimuthal angle $\phi$ of the two leptons and two candidate jets.
We constrain the dilepton invariant mass to $91.2 \pm 2.5$ GeV and the vector sum of $p_T$ to $0.0 \pm 7.0$ GeV.

Subsequently, we remove events with high kinematic fit $\chi^2$ values to reduce instrumental background.
As a result of the fit, the dijet mass resonance in Higgs events is more prominent, which translates directly
to a 6-11\% increase in Higgs sensitivity.

\section{Results}

\begin{figure}
\centerline{\includegraphics[width=0.7\textwidth]{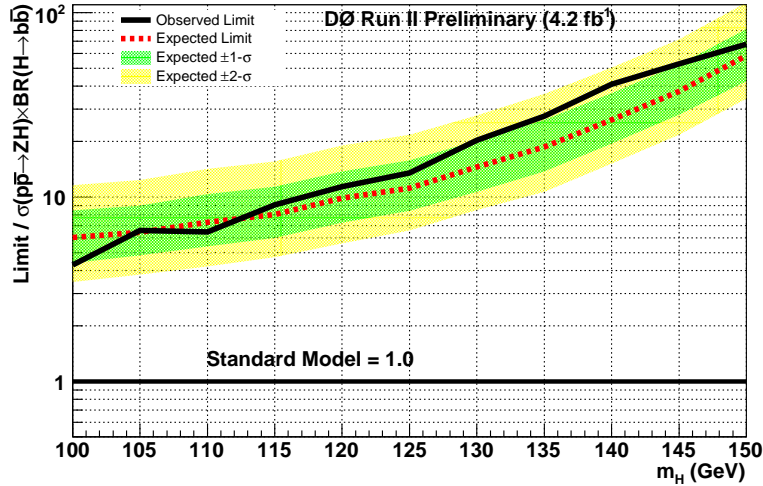}}
\caption{Expected and observed limits on $ZH$ production, 
         expressed as ratios to the standard model expectations.}\label{fig:limits}
\end{figure}

We use boosted decision trees (BDT) \cite{bdt} to combine the discrimination power of several kinematical variables:
the dijet mass and $p_T$, the dilepton $p_T$ and colinearity, and many others.
No evidence for $ZH$ production is seen, so we compute upper cross section limits based on the shape
of the BDT output, using a modified frequentist approach \cite{semifreq, collie}.  The leading sources of systematic uncertainty
are the $Z+$heavy-flavor cross section (20\%), the jet energy scale (10\%), and b-tagging efficiencies (10\%).
Assuming a Higgs mass of 115 GeV, we exclude $ZH$ production above 9.1 times the standard model expectation
at 95\% confidence level.  Limits assuming other Higgs masses are shown in Fig. \ref{fig:limits}.

Upon comparison to previous results \cite{previous}, our limits have improved roughly 12\% beyond what is
expected with more data. This is due to our use of more optimal selection criteria and more
sophisticated analysis techniques. We are currently investigating further improvements to the analysis, including
the use of matrix-element discriminants, improved b-tagging, and further optimization of our
multivariate discriminant.

\end{document}